\documentclass[10pt,journal]{IEEEtran}
\usepackage{color, soul, framed}
\usepackage{enumerate}
\usepackage{amsfonts}
\usepackage{graphicx}
\usepackage{color}
\usepackage{amsmath,amsfonts,amssymb,amsthm,epsfig,epstopdf,array}
\usepackage{url,textcomp}
\usepackage{mdwmath}
\usepackage{mdwtab}
\usepackage{times}
\usepackage{authblk}
\usepackage{cite}
\usepackage{amsbsy}
\usepackage{mathtools}
\usepackage{subfigure}
\usepackage{algorithmic}
\usepackage{stfloats}
\usepackage[misc]{ifsym}

\DeclareMathAlphabet{\mathpzc}{OT1}{pzc}{m}{it}
\newtheorem{theorem}{Theorem}

\usepackage{bicaption}
\begin{document}
\title{Outage Performance Analysis of HARQ-Aided Multi-RIS Systems} \author{\IEEEauthorblockN{
Qi~Cao\IEEEauthorrefmark{1},
Huan~Zhang\IEEEauthorrefmark{2}, Zheng~Shi$^{\textrm{\Letter}}$\IEEEauthorrefmark{3},
Hong~Wang\IEEEauthorrefmark{4},
Yaru~Fu\IEEEauthorrefmark{5},
Guanghua~Yang\IEEEauthorrefmark{3}, and
Shaodan~Ma\IEEEauthorrefmark{2}\\
\vspace{-1em}
\IEEEauthorrefmark{1}School of Electronics and Information Technology, Sun Yat-Sen University, China\\
\IEEEauthorrefmark{2}Department of Electrical and Computer Engineering, University of Macau, China}\\
\IEEEauthorrefmark{3}School of Intelligent Systems Science and Engineering, Jinan University, Zhuhai, China\\
\IEEEauthorrefmark{4}School of Communication and Information Engineering, Nanjing University of Posts and Telecommunications, Nanjing, China\\
\IEEEauthorrefmark{5}School of Science and Technology, The Open University of Hong Kong, Hong Kong, China\\
\vspace{-1.5em}}

\maketitle
\begin{abstract}
Reconfigurable intelligent surface (RIS) has recently attracted a spurt of interest due to its innate advantages over Massive MIMO on power consumption.  In this paper, we study the outage performance of multi-RIS system with the help of hybrid automatic repeat request (HARQ) to improve the RIS system reliability, where the destination received channels are modeled by Rician fading and the  phase shift setting only depends on the line-of-sight (LoS) component. Both the exact and asymptotic outage probabilities under Type-I HARQ and HARQ with chase combining (HARQ-CC) schemes are derived. Particulary, the tractable asymptotic results empower us to derive meaningful insights for HARQ-aided multi-RIS system. On the one hand, we find that both the Type-I and the HARQ-CC schemes can achieve full diversity that is equal to the maximal number of HARQ rounds. On the other hand, the closed-form expression of the optimal phase shift setting with respect to outage probability minimization is obtained. The optimal solution indicates that the reflecting link direction should be consistent with direct link LoS component. Finally, the analytical results are validated by Monte-Carlo simulations.


\end{abstract}
\IEEEpeerreviewmaketitle
\hyphenation{HARQ}
\section{Introduction}\label{sec:int}

\IEEEPARstart The explosive growth of wireless data traffic worldwide has accelerated the energy consumption in wireless networks. Therefore, spectrum and energy efficient designs have caused extensive discussion. Reconfigurable Intelligent Surfaces (RIS), as a recently effective hardware technology to improve the spectrum efficiency with lower power consumption, has been regarded as a potential technology for beyond 5G\cite{hu2018beyond,wu2020towards,basar2020reconfigurable}. Specifically, RIS is an artificial metamaterial, which is composed of a large number of low-cost, passive and reflective elements with reconfigurable parameters. It has the ability to intelligently adjust the reflecting signal phase to achieve certain communication objectives, e.g., increasing the received signal energy, expanding the coverage region, alleviating interference, etc. \cite{ElMossallamy2020}.


 Due to the advantages of smart adjustment and low power consumption, RIS has been an active research area recently in various aspects to improve the performance of wireless communication systems \cite{wu2019intelligent, kammoun2020asymptotic,boulogeorgos2020performance,li2020weighted,Atapattu2020reconfigurable}.
In \cite{wu2019intelligent}, the authors proposed a joint transmit precoding and passive phase shift design for RIS-aided systems to minimize the transmit power. However, the assumption of prefect channel state information (CSI) in \cite{wu2019intelligent} is impractical due to the high signal processing complexity. To tackle the challenges, a deterministic approximation for signal-to-noise-interference-plus-noise-ratio (SINR) was derived in \cite{kammoun2020asymptotic}, which eases the phase shift matrix design with statistics CSI. To investigate the efficiency of RIS-aided wireless systems, the authors in \cite{boulogeorgos2020performance} studied the performance for Rayleigh channels and demonstrated its superior performance over amplify-and-forward (AF) relaying scheme in terms of average symbol error rate (SER), ergodic capacity (EC), etc.. To further enhance the RIS-aided system performance, the multi-RIS assisted systems were studied in \cite{li2020weighted}, and an effective method was proposed to maximize the weighted sum rate.  Moreover, the performance of RIS assisted two-way systems over Rayleigh fading channels was investigated in \cite{Atapattu2020reconfigurable}, and the exact expressions for OP and EC were derived in closed-form.


Although the wireless propagation environment can be ameliorated with the help of RIS, the problem of decoding outage always exists, such as long-distance communication and severe fading \cite{ji2018ultra}. To enhance the reliability of wireless communication systems, hybrid automatic repeat request (HARQ) is proved to be a feasible way. The advantage of HARQ is attributed to the retransmissions of the received packets that fail to be decoded with forward error correction (FEC) \cite{shi2017asymptotic, shi2018energy}. In general, the HARQ techniques can be  classified into three types based on the encoding and decoding operations at the transceivers, i.e., Type I HARQ, HARQ with chase combining (HARQ-CC) and HARQ with incremental redundancy (HARQ-IR)\cite{shi2019achievable}. Specifically, in Type I HARQ, the erroneously received packets are directly discarded and each received packet is independently decoded. In HARQ-CC, the erroneous packets are stored and then combined with the retransmitted packets by using maximal ratio combining (MRC). In HARQ-IR, code combination is used to concatenate the currently received packet and failed packets for joint decoding\cite{Caire2001the}. The authors in \cite{Wang2020outage} investigated the outage performance of  non-orthogonal multiple access (NOMA)-aided small cell network with Type-I HARQ, and the  closed-form expression for OP was derived by using  stochastic geometry. In \cite{chelli2018throughput}, the authors considered double Rayleigh channels to study the performance of HARQ-CC communication systems. Furthermore, the outage probability for HARQ-IR under time-correlated rayleigh fading channels was investigated in \cite{Shi2015analysis}.

In order to provide reliable data transmission for RIS systems, this paper proposes the HARQ-aided multi-RIS systems under two HARQ schemes, i.e., Type-I HARQ and HARQ-CC. We focus on the Type-I HARQ and the HARQ-CC schemes because of their low complexity by comparing to the HARQ-IR scheme. In contrast to \cite{kammoun2020asymptotic,boulogeorgos2020performance,li2020weighted,Atapattu2020reconfigurable} that assumed no direct link, this paper considers a more practical scenario that the destination node can receive the signals from both direct and reflecting links. This assumption is more applicable to outdoor communication scenarios. Besides, the assumption of Rician fading channels is more practical \cite{xiong1997performance}. To investigate the outage performance of HARQ-assisted multi-RIS systems, we derive closed-form expressions for the exact and asymptotic outage probabilities of the two HARQ schemes. On the basis of the exact results, the asymptotic outage probabilities are derived for the sake of analytical tractability. These asymptotic results enable us to capture more meaningful insights. To be specific, we derive the diversity order for each HARQ scheme. It is found that both the Type-I and the HARQ-CC schemes can achieve the full diversity and the diversity order is equal to the maximal number of HARQ rounds. To further enhance the outage performance, we formulate an outage minimization problem with respect to the phase shift. The optimal solution can be obtained with the outage expressions. The results show that the optimal phase setting should make the reflecting link direction consistent with direct LoS link. To the best of our knowledge, this is the first work that investigates the outage performance for HARQ-aided RIS systems.

%
The rest of this paper is organized as follows. Section II presents the system model of HARQ-aided multi-RIS system. The closed-form expressions for exact and asymptotic outage probabilities are derived in Section III and IV, respectively. Moreover, some meaningful insights are gained in Section IV. Simulation and numerical results are provided in Section V and Section VI concludes the paper.

\section{System Model}\label{sec:sys_mod}
\begin{figure}[!t]
\centering
\includegraphics[width=3in]{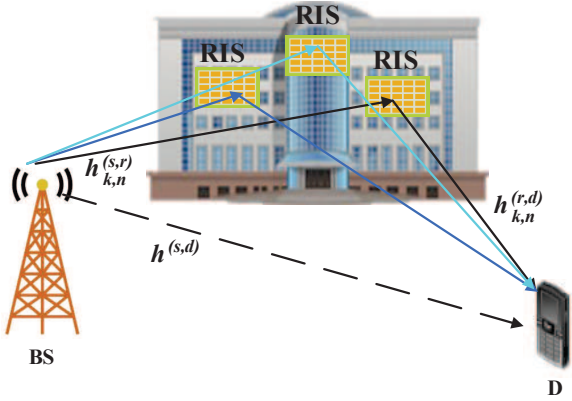}
\caption{The $l$th transmission round for HARQ-aided Multi-RIS system.}
\end{figure}
As shown in Fig. 1, we consider a HARQ-aided RIS communication system that includes a base-station (S), a destination node (D) and $K$ multiple RISs. Both S and D are equipped with a single antenna, each RIS composed of $N_k, k=1,\cdots,K$ passive reflecting elements mounted on the wall of the high-rise building to assist the communication between S and D. As opposed to \cite{kammoun2020asymptotic,boulogeorgos2020performance,li2020weighted,Atapattu2020reconfigurable}, we assume D can receive the signal from both direct channel ``S-D'' and
cascaded reflecting channel ``S-RIS-D''. This assumption is more practical in outdoor communication scenarios. In addition, similarly to the existing works \cite{hu2018user,li2020weighted}, the signal reflected between the multi-RIS for more times is negligible and can be ignored for analytical simplicity.

\subsection{Channel Model}
 Let ${h^{(s,d)}}$, $h_{k,n}^{(s,r)}$, and ${h_{k,n}^{(r,d)}}$ denote the channel of S-$(k,n)$th RIS element, $(k,n)$th RIS element-D, and S-D links, respectively, where $(k,n)$  represents the  $n$th reflection unit of the $k$th RIS, $n=1,\cdots,N_k$. In particular, the channel between S and $(k,n)$th RIS element have a highly possibility to have line-of-sight (LoS) communication link due to the high altitude of BS and RIS. Hence, the channel between S and $(k,n)$th reflecting element can be modeled as
 \begin{equation}\label{eqn:hsr}
h_{k,n}^{(s,r)} = \sqrt {{\beta}_k^{(s,r)}} \bar h_{k,n}^{(s,r)},
\end{equation}
 where ${\alpha _k}^{(s,r)}$ and  $ \bar{h}_{k,n}^{(s,r)}$ represent the distance dependent path-loss and the deterministic LoS components with $|\bar{h}_{k,n}^{(s,r)}|=1$ normalization, respectively.

Different from $h_{k,n}^{(s,r)}$, the channels for modeling $(k,n)$th RIS element to D, or S to D links always consist of components of LoS and non-LoS (NLoS). Since the received signal at D is constituted by a lot of reflection and scattering components from building and ground. Rician fading channels are assumed to be ${h^{(s,d)}}$  and $h_{k,n}^{(r,d)}$ for the links of ``S-D'' and ``RIS-D'', respectively. More precisely, $h^{(s,d)}$ and $h_{k,n}^{(r,d)}$ can be modeled as
 \begin{equation}\label{eqn:hsd}
{h^{(s,d)}} = \sqrt {{\beta ^{(s,d)}}} \big(\sqrt {\frac{{{\kappa ^{(s,d)}}}}{{{\kappa ^{(s,d)}} + 1}}} {{\bar h}^{s,d}} + \sqrt {\frac{1}{{{\kappa ^{(s,d)}} + 1}}} {\widetilde h^{s,d}} \big),
\end{equation}

\begin{equation}\label{eqn:hrd}
h_{k,n}^{(r,d)} = \sqrt {{\beta _k}^{(r,d)}}\big( \sqrt {\frac{{{\kappa _k}^{(r,d)}}}{{{\kappa _k}^{(r,d)} + 1}}} \bar h_{k,n}^{(r,d)} + \sqrt {\frac{1}{{{\kappa _k}^{(r,d)} + 1}}} \widetilde h_{k,n}^{(r,d)}\big),
\end{equation}
 where ${\beta^{(s,d)}},{\beta _k}^{(r,d)}$  represent the path loss depending on the distance, ${{\bar h}^{s,d}},\bar h_{k,n}^{(r,d)}$  denote the deterministic LoS component normalized by $1$, ${\widetilde h^{s,d}},\widetilde h_{k,n}^{(r,d)}\sim\mathcal{CN}(0,1)$  correspond to normalized NLoS components with zero mean and unit variance, and  ${\kappa ^{(s,d)}},{\kappa _k}^{(r,d)} \ge 0$ denote the Rician factors, respectively. Furthermore, it is noted that the LoS component and distance remain unchanged during HARQ transmissions.
 \subsection{HARQ Protocol}
 In order to improve the transmission reliability of RIS-assisted wireless communication system, two types of HARQ schemes, i.e. Type-I of HARQ and HARQ-CC,  are considered in this paper. We assume that the number of HARQ rounds for the two HARQ schemes is limited to $L$. Hence, the signal received at D in the $l$th HARQ round is given by

 \begin{equation}\label{eqn:received signal}
{y_l}= \sqrt {{P}} \underbrace {{\left[{h^{(s,d)}} +\sum\limits_{k=1}^K\sum\limits_{n=1}^N {{h_{k,n}^{(r,d)}{\alpha_{k,n}e^{j{\theta _{_{k,n}}}}}h_{k,n}^{(s,r)}}}  \right]}}\limits_{ \buildrel \Delta \over = {h_l}} {x_l} + {z},
\end{equation}
 where $P$, $x$ and $z\sim\mathcal{CN}(0,{\sigma ^2})$ denote the transmit power, transmitted symbol and additive white Gaussian noise (AWGN), respectively, $ l=1,\cdots,L$, $\alpha_{k,n}e^{j\theta_{k,n}}$ stands for the response of the $(k,n)$th RIS reflection element, $\alpha_{k,n}\in (0,1]$ and $\theta_{k,n}\in [0,2\pi]$ represents the amplitude refection coefficient and phase shift after reflection, respectively. ${h_l}$ denotes the equivalent
  channel  between the BS and node D. For analytical simplicity, we assume perfect reflection at RIS, i.e.,  $\alpha_{k,n}=1$ , and  the design of the phase shift coefficient $\theta_{k,n}$ depends only on the LoS component and is irrelevant to instantaneous NLoS components.

 \subsubsection{Type I HARQ}
 In Type-I HARQ, the reciever D attempts to decode the message based on merely the data packets that just received. To be specific, the erroneously received packets are discarded and each retransmitted packet is decoded independently. Then, the accumulated mutual information obtained by Type I HARQ after $L$ HARQ rounds can be expressed as
 \begin{equation}\label{Type_I}
\mathcal{I}^{Type-I}=\mathop {\max }\limits_{l=1,\cdots,L}\mathrm{log}_{2}(1+ {\rho}{|h_l|}^2).
 \end{equation}
 \subsubsection{HARQ-CC}
 In HARQ-CC, the previously failed packets are stored and combined with the currently received packets. Accordingly, the accumulated mutual information of the HARQ-CC after $L$ HARQ rounds is given by
  \begin{equation}\label{Type_CC}
\mathcal{I}^{CC}=\mathrm{log}_{2}(1+ \sum\limits_{l=1}^L{\rho}{|h_l|}^2),
 \end{equation}
 where $\rho=P/\sigma^2$ stands for the average signal-to-noise ratio (SNR).

 \section{Outage Probability Analysis}
 To investigate transmission reliability of HARQ-aided multi-RIS systems with respect to the above two schemes, the outage probability (OP) is the most important performance metric that is defined as the probability of the accumulated mutual information at the destination is less than the transmission rate  $R$ \cite{shi2017asymptotic,yang2020asymptotic}. Accordingly, the OP for HARQ can be written as
 \begin{equation}\label{eqn:Pout_I}
P_{out} = \Pr(\mathcal{I} < \mathcal{R}),
\end{equation}
where $\mathcal{I}\in \{\mathcal{I}^{Type-I}, \mathcal{I}^{CC}\}$.
The outage analyses for the aforementioned two types of HARQ schemes will be undertaken individually in the following subsections.

 \subsection{OP for Type-I HARQ}
 To proceed with the Type-I OP analysis, it is necessary to determine the distribution of ${\left| h_l \right|^2}$ , as given by the following theorem.
 \begin{theorem}\label{Theorem_I}
 The  cumulative distribution function(CDF) of the channel gain for the $l$th HARQ round is given by \cite{zhang2019analysis}
 \begin{equation}\label{eqn:PDF_h_l}
{F_{{\left| h_l \right|^2}}}(x) = {e^{ - \frac{\Psi_{gLoS}(\boldsymbol{\theta} )}{\Psi_{gNLoS}}}}\sum\limits_{i = 0}^\infty  {\frac{{{{(\frac{\Psi_{gLoS}(\boldsymbol{\theta} )}{\Psi_{gNLoS}})}^i}}}{{i!}}} \frac{{\gamma (1 + i,\frac{x}{\Psi_{gNLoS}})}}{{\Gamma (1 + i)}},
\end{equation}
 where $\boldsymbol{\theta}$  denotes the set of $(\theta_{k,n}),k\in{1,\cdots,K},n\in{1,\cdots,N_k}$, the expressions for $\Psi_{gLoS}(\boldsymbol{\theta} )$ and  $\Psi_{gNLoS}$ are given by  (\ref{eqn:glos}) and (\ref{eqn:gnlos}), as shown at the top of next page, which represent the power of the LoS and NLoS component of the equivalent channel,  respectively. $\gamma (u,v) \buildrel \Delta \over = \int_0^v {{x^{u - 1}}{e^{ - x}}} dx$ is the incomplete gamma function \cite[Eq. (8.350.1)]{gradshteyn2014table}.
\begin{figure*}[!t]
\begin{align}\label{eqn:glos}
\Psi_{gLoS}(\boldsymbol{\theta} )\buildrel \Delta \over = \big|\sqrt {\frac{{{\beta ^{(s,d)}}{\kappa ^{(s,d)}}}}{{{\kappa ^{(s,d)}} + 1}}} {{\bar h}^{s,d}} +
\sum\limits_{k \in K,n \in {N_k}} {\sqrt {\frac{{{\beta _k}^{(r,d)}{\beta _k}^{(s,r)}{\kappa _k}^{(r,d)}}}{{{\kappa _k}^{(r,d)} + 1}}} } \bar h_{k,n}^{(r,d)}{e^{j{\theta _{_{k,n}}}}}\bar h_{k,n}^{(s,r)}{ \big|^2}
\end{align}
\hrulefill
\begin{equation}\label{eqn:gnlos}
\Psi_{gNLoS} \buildrel \Delta \over = \frac{{{\beta ^{(s,d)}}}}{{{\kappa ^{(s,d)}} + 1}} + \sum\limits_{k \in K} {\frac{{{N_k}{\beta _k}^{(r,d)}{\beta _k}^{(s,r)}}}{{{\kappa _k}^{(r,d)} + 1}}}
\end{equation}
\hrulefill
\end{figure*}
  \end{theorem}
As a consequence, substituting (\ref{Type_I}) and (\ref{eqn:PDF_h_l}) into (\ref{eqn:Pout_I}) yields a
closed-form expression of OP for Type-I HARQ scheme as
 \begin{equation}\label{eqn:Pout_I}
{P_{out}^{Type-I}} = \big[{e^{ - \frac{\Psi_{gLoS}(\boldsymbol{\theta} )}{\Psi_{gNLoS}}}}\sum\limits_{i = 0}^\infty  {\frac{{{{(\frac{\Psi_{gLoS}(\boldsymbol{\theta} )}{\Psi_{gNLoS}})}^i}}}{{i!}}} \frac{{\gamma (1 + i,\frac{\psi}{\Psi_{gNLoS}})}}{{\Gamma (1 + i)}}\big]^L,
\end{equation}
where $\psi=\frac{{{2^R} - 1}}{\rho}$.

\subsection{OP for HARQ-CC}
By observing eq. (\ref{Type_CC}), the distribution of $G\buildrel \Delta \over = \sum\nolimits_{l=1}^L{\rho}{|h_l|}^2$ acts as an essential role in deriving the OP of HARQ-CC. However, there is no off-the-shelf result due to the difficulty of tackling the sum of multiple random variables, which considerably challenges the subsequent outage analysis. With the help of the density function for the non-central $\chi^2$ distribution, we can derive the CDF expression of $G$ which is given by the following theorem.
\begin{theorem}\label{Theorem_II}
 The CDF of $G$ can be obtained as an infinite summation such that
\begin{align}\label{eqn:CDF_CC}
F_G(x)=   {e^{ - {\xi}}}\sum\limits_{i = 0}^\infty  {\frac{{{{(\xi)}^i}}}{{i!}}} \frac{{\gamma (L + i,\frac{x}{\Psi_{gNLoS}})}}{{\Gamma (L + i)}},
\end{align}
where $\xi=L\Psi_{gLoS}(\boldsymbol{\theta} )/{\Psi_{gNLoS}}$.
\begin{proof}
Please see Appendix \ref{app:proof_Theorem_II}.
\end{proof}
  \end{theorem}
According to (\ref{eqn:CDF_CC}), the OP for multi-RIS system under HARQ-CC scheme is given by
 \begin{equation}\label{Pout_CC}
{P_{out}^{CC}} =  {e^{ - \xi}}\sum\limits_{i = 0}^\infty  {\frac{{(\xi )}^i}{i!}\frac{{\gamma (L + i,\frac{\psi}{\Psi_{gNLoS}})}}{{\Gamma (L + i)}}}.
\end{equation}
Based on (\ref{eqn:Pout_I}) and (\ref{Pout_CC}), it is easily found that the closed-form outage probabilities for each HARQ scheme have many terms in common and
suitable phase shift setting have positive impacts. However, the outage probabilities in (\ref{eqn:Pout_I}) and (\ref{Pout_CC}) involved incomplete gamma function and hardly to extract insightful results. In next section, we will obtain meaningful insights by virtue of asymptotic analyses.
%
%
%
%

\section{Asymptotic Analysis and Discussions}\label{sec:opt_div}
In order to capture some useful insights for two scheme and minimize the outage probabilities by optimizing the phase shifts of multi-IRS, we consider high SNR regime, i.e., $\rho\rightarrow \infty$, to derive the asymptotic outage probabilities.
\subsection{Asymptotic Analysis}
By using (12) and exchanging the order of summations and multiplications the Type-I HARQ OP can be rewritten as
\begin{align}\label{eqn:pout_I_dio}
&{P_{out}^{Type-I}}\notag \\
 &= \sum\limits_{{i_1}, \cdot  \cdot  \cdot {i_L} = 0}^\infty  {{e^{ - L\frac{{\Psi_{gLoS}(\boldsymbol{\theta} )}}{{\Psi_{gNLoS}}}}}} \prod\limits_{l = 1}^L {\frac{{{(\frac{{\Psi^2_{gLoS}(\boldsymbol{\theta} )}}{{\Psi_{gNLoS}}})}^{i_l}}}{i_l}} \frac{{\gamma (1 + {i_l},\frac{\psi}{{\Psi_{gNLoS}}})}}{\Gamma (1 + {i_l})}.
\end{align}
By adopting the series representations of the incomplete gamma function  $\gamma (\alpha,x) = \sum\limits_{{j} = 0}^\infty  {\frac{{{{( - 1)}^{{j}}}}}{{{j}!(\alpha+j)}}} {(x)^{\alpha+ j}}$\cite[Eq. (8.351.1)]{gradshteyn2014table}, the OP for Type-I HARQ scheme is asymptotic to
\begin{align}\label{eqn:pout_I_asy}
{P_{out,\infty}^{Type-I}}= {e^{ - L\frac{{\Psi_{gLoS}(\boldsymbol{\theta})}}{{\Psi_{gNLoS}}}}}{\left( {\frac{\psi}{{\Psi_{gNLoS}}}} \right)^L} + \mathcal{O}\left( {{\rho^{-L}}} \right),
\end{align}
where $\mathcal{O}(\cdot)$ denotes higher order terms.

In an analogous way, we can also derive the asymptotic OP for HARQ-CC scheme as
\begin{align}\label{eqn:pout_CC_asy}
{P_{out,\infty}^{CC}}= \frac{{{e^{ - L\frac{{\Psi_{gLoS}(\boldsymbol{\theta})}}{\Psi_{gNLoS}}}}}}{{\Gamma (L + 1)}}{\left( {\frac{\psi}{\Psi_{gNLoS}}} \right)^L} + \mathcal{O}\left( \rho^{-L}\right).
\end{align}
\subsection{Discussions for Asymptotic Results}
As suggested by \cite{yang2020asymptotic}, in the high SNR regime, i.e., $\rho\rightarrow \infty$, the asymptotic OP can be expressed as
\begin{align}\label{re_asy}
P_{out,\infty}=\mathcal{S}_{LoS,NLoS}(\mathcal{C}(R)\rho)^{-d}+\mathcal{O}(\rho^{-d}),
\end{align}
where $\mathcal{S}_{LoS,NLoS}$ quantifies the impact of phase shift, LoS and NLoS component, $\mathcal{C}(R)$ is the modulation and coding gain, and $d$ stands for the diversity order.
\subsubsection{ Impact of Phase Shift}
By substituting (\ref{eqn:pout_I_asy}) and (\ref{eqn:pout_CC_asy}) into (\ref{re_asy}), we can obtain
\begin{equation}\label{Impact_Phase}
\mathcal{S}^{Type-I}_{LoS,NLoS}=\mathcal{S}^{CC}_{LoS,NLoS}={e^{ - L\frac{{\Psi_{gLoS}(\boldsymbol{\theta})}}{{\Psi_{gNLoS}}}}}\Psi_{gNLoS}^{-l}.
\end{equation}
From (\ref{Impact_Phase}), we can easily find that the phase shift affects asymptotic outage probabilities only through the term of $\Psi_{gLoS}(\boldsymbol{\theta})$. Since both Type-I HARQ and HARQ-CC are monotonically decreasing with respect to $\Psi_{gLoS}(\boldsymbol{\theta})$, the OP minimization problem with respect to $\boldsymbol{\theta}$ amounts to
\begin{equation}
\begin{aligned}
 \label{eq_optimal}
\mathcal{P}\,:\,&\max\limits_{\boldsymbol{\theta}}\ \Psi_{gLoS}(\boldsymbol{\theta}) \\
& \begin{array}{r@{\quad}r@{}l@{\quad}l}
\mathrm{s.t.}
&\,&\boldsymbol{\theta} \buildrel \Delta \over=(\theta_{k,n})_{k\in{1,\cdots,K},n\in{1,\cdots,N_k}}.\\
&\,&0\leq\theta_{k,n}\leq 2\pi.
\end{array}
\end{aligned}
\end{equation}
Due to the similar form of the optimization problem to \cite{zhang2019analysis}, the optimal solution of the phase shift $\theta^*$ can be found in \cite[Theorem 2]{zhang2019analysis}, and the results are omitted here to save space.
\subsubsection{Modulation and Coding Gain}
The modulation and coding gain $\mathcal{C}(R)$ quantifies the amount of the SNR reduction required to reach the same outage probability \cite{yang2020asymptotic}. By substituting (\ref{eqn:pout_I_asy}) and (\ref{eqn:pout_CC_asy}) into (\ref{re_asy}), we can derive
\begin{equation}\label{MCG_typeI}
\mathcal{C}^{Type-I}(R)=(2^R-1)^{-1},
\end{equation}
and
\begin{equation}\label{MCG_CC}
\mathcal{C}^{CC}(R)=(\Gamma (L + 1))^{\frac{1}{L}}({2^R-1})^{-1}.
\end{equation}
It is easily found from (\ref{MCG_typeI}) and (\ref{MCG_CC}) that the modulation and coding gain for HARQ-CC always larger than Type-I HARQ for $L\geq 2$. Hence, by comparing with Type-I HARQ scheme, HARQ-CC scheme can achieve the same outage requirement with lower SNR.
\subsubsection{Diversity Order}
the diversity order is a fundamental asymptotic reliability metric to characterize the degree of freedom for a communication system. More specifically, the diversity order describes how the OP scales with the transmit SNR on a log-log scale. By substituting (\ref{eqn:pout_I_asy}) and (\ref{eqn:pout_CC_asy}) into (\ref{re_asy}), we can easily derive the diversity order for each HARQ scheme as
\begin{align}
d^{Type-I}=d^{CC}=L.
\end{align}
 In contrast to \cite{kudathanthirige2020performance}, the diversity order in our paper is irrelevant to the number of RISs and reflecting elements which can be explained as follows. With statistic CSI at S, the phase shift is fixed for each HARQ round. Accordingly, $h_l$ can be regarded as an aggregated channel composing of direct channel and multi-reflecting channel. Moreover, it is worthwhile to note that both the Type-I and HARQ-CC schemes can achieve the full diversity and the diversity order is equal to the number of HARQ rounds.

\section{Numerical Results}\label{sec:num}
\begin{figure}[!t]
\centering
\includegraphics[width=3.5in]{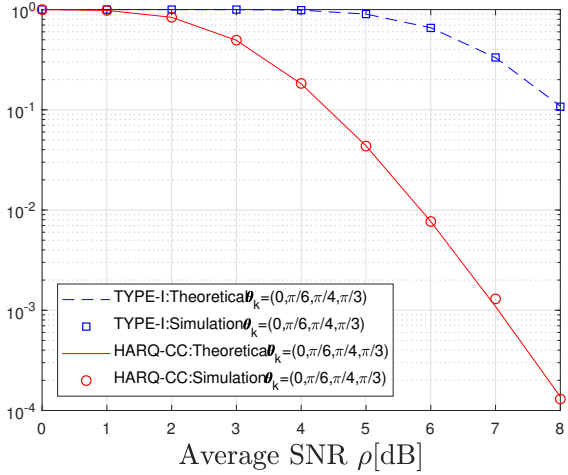}
\caption{The outage probability versus the average SNR with $L=4$.}
\end{figure}

\begin{figure}[!t]
\centering
\includegraphics[width=3.5in]{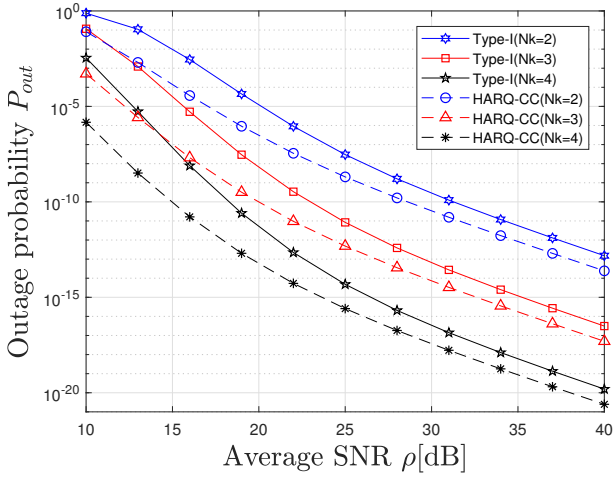}
\caption{The outage probability versus the average SNR under different  reflecting elements number  with $L=3$ and { ${\boldsymbol{\theta}}_{k}=(0,\frac{\pi}{6},\frac{\pi}{4},\frac{\pi}{3})$}.}
\end{figure}
In this section, numerical results are presented to verify our analytical results.  Unless otherwise indicated, the system parameters are set as follows,  $K = 3$, ${N_k} = 4$, $R = 4 \mathrm{bps/Hz}$, $\kappa^{(s,d)} = -5\,{\mathrm{dB}}$, $\kappa^{(r,d)} = 0.4\,{\mathrm{dB}}$, $L=4$, and each RIS has the same phase shift setting vector ${\boldsymbol{\theta}}_{k}$. The distance dependent path loss is modeled as $\beta=({d_i}/{d_0})^{-\alpha}$, where $d_i$ denotes the communication distance,  $d_0=20$m stands for a reference distance. Herein, we assume $d_{sd}=70m$, $d_{sr_k}=50m$, $d_{{r_k}d}=40m$. The path-loss exponents for each link are set as $\alpha_{sd}=2.5$, $\alpha_{sr_k}=2$, $\alpha_{r_kd}=2.2$, where $k\in\{1,\cdots,K\}$.
Although eqs. (\ref{eqn:Pout_I}) and (\ref{Pout_CC})
provide closed-form expressions for outage probabilities, the sum of infinite series imposes significant challenges on numerical implementation. To enable the computation of (\ref{eqn:Pout_I}) and (\ref{Pout_CC}), we truncate (\ref{eqn:Pout_I}) and (\ref{Pout_CC}) to finite summations up to the truncation order $i$.
By comparing the finite truncated expressions and simulation, we find that the approximate $i=50$ can meet the accuracy requirement of $10^{-3}$ under the aforementioned parameter settings. Hence, $i=50$ is set in the sequel.

Figs. 2 and 3 depict the OP versus the average SNR under different phase shift parameter and reflecting elements number, respectively. It can be observed from
both figures that the outage probability decreases as the average SNR increases. Moreover, and the exact and simulation results are in perfect agreement for the two HARQ schemes, which confirms the correctness of the OP analyses. By comparing the two HARQ schemes, it is observed that the HARQ-CC scheme always  outperforms Type-I HARQ, this is because the HARQ-CC scheme can combine the retransmit signals to improve the received SNR.  In addition, for the fixed phase shift $\boldsymbol{\theta}$ in Fig. 3, the OP can be enhanced by utilizing more reflecting elements on each RIS.

Fig. 4 presents the exact and asymptotic outage versus SNR  with different HARQ rounds according to Sections III and IV-A. One can observe that the asymptotic curves tightly approximate the exact curves with the increase of SNR, which validates the asymptotic results. Furthermore, the slopes of all outage curves are identical and coincide with the maximal number of HARQ rounds. This result is consistent with the diversity order analysis in Section VI-B.

The impact of the phase shift adjustment on the OP is presented in Fig. 5.  For illustration, we assume that each IRS equips with $N_k=4$ reflecting elements
as an example, wherein the labels ``optimal", ``fixed" and ``random" refer to the  optimal $\theta_{k,n}^\ast$, constant value $\theta_{k,n}=\frac{\pi}{3}$, and uniform distribution selection $\theta_{k,n}\sim U[0,2\pi]$, respectively. It can be observed from Fig. 5 that the optimal phase shift $\theta_{k,n}^\ast$ performs the best among the three phase setting strategies in terms of the OP. 
\begin{figure}[!t]
\centering
\includegraphics[width=3.5in]{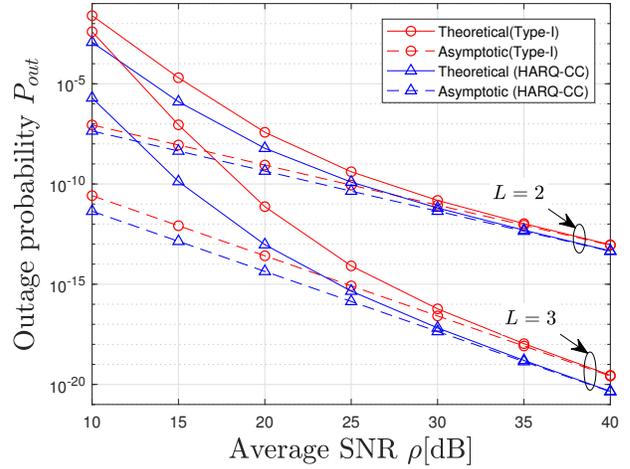}
\caption{The outage probability versus the average SNR under different HARQ rounds with $N_k=4$ and ${\boldsymbol{\theta}}_{k}=(0,\frac{\pi}{6},\frac{\pi}{4},\frac{\pi}{3})$.}
\end{figure}

\begin{figure}[!htbp]
\centering
\includegraphics[width=3.5in]{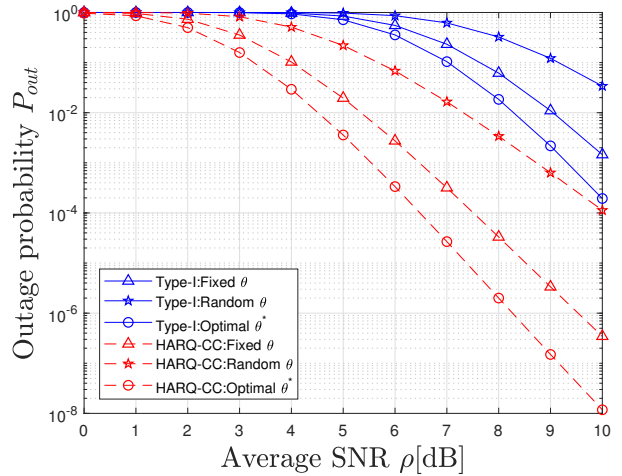}
\caption{The outage probability versus the average SNR under different phase shift setting.}
\end{figure}

\section{Conclusions}\label{sec:con}
This paper has investigated the outage performance of multi-RIS wireless communication systems, where the Type-I HARQ and HARQ-CC schemes have been adopted to enhance the communication reliability. The exact and asymptotic closed-form expressions for the outage probabilities have been derived and validated by simulations. Numerical results have illustrated that the outage performance of multi-RIS system can be improved with the increase of SNR and the number of HARQ rounds. Besides, it has been proved that the HARQ-CC scheme performs better than Type-I HARQ owing to the recycle of previously received packets. According to the expressions for asymptotic outage probabilities, we have found that full diversity can be achieved for both schemes and equals to the maximal number of HARQ rounds. Finally, the OP minimization  with respect to phase shift has been considered. We have drawn the conclusion that the optimal phase shift setting for reflecting link should be consistent with direct link LoS. To the best of our knowledge, this is the first work to deals with the outage performance of HARQ-aided multi-IRS systems.

%
\appendices
\section{Proof of Theorem \ref{Theorem_II}} \label{app:proof_Theorem_II}
It is readily found that $ {h_l}$ is  a  complex Gaussian random variables with means $\Psi_{gLoS}(\boldsymbol{\theta} )$ and variance $\Psi_{gNLoS}$, i.e. ${h_l}\sim\mathcal{CN}(\Psi_{gLoS}(\boldsymbol{\theta} ),\Psi_{gNLoS})$. Thus, the sum of squares for  complex Gaussian random variables$\frac{G}{\Psi_{gNLoS}/2}=\frac{1}{\Psi_{gNLoS}/2}\sum\limits_{l=1}^L{h_l}^2$ is a non-central $\chi^2$ distribution  with $2L$ degrees of freedom. With the help of \cite[Theorem 1.3.4]{muirhead2009aspects}, we can derive the probability density function (PDF) of $Z=\frac{G}{\Psi_{gNLoS}/2}$ as
\begin{equation}\label{eqn:func_RV}
f_Z(z) = {e^{ - \xi}}{}_0{F_1}(L;\frac{1}{2}\xi z)\frac{1}{{{2^L}\Gamma (L)}}{e^{ - \frac{z}{2}}}{z^{L - 1}},
\end{equation}
 where $\xi=L \Psi_{gLoS}(\boldsymbol{\theta} )/\Psi_{gNLoS}$ and ${}_0{F_1}(;)$ represents the generalized hypergeometric series \cite[Eq. (9.14.1)]{gradshteyn2014table}. By putting ${}_0{F_1}(L;\frac{1}{2}\xi z)=\sum\limits_{i = 0}^\infty  {\frac{{(\frac{1}{2}\xi z)}^i}{{{(L)}_i}i!}} $ into (\ref{eqn:func_RV}), where ${{(L)}_i}=L(L+1)\cdots(L+i-1)$,  and utilizing  \cite[Eq. (3.351.1)]{gradshteyn2014table}, we can derive the CDF of $Z$ as
\begin{align}\label{eqn:CDF_Z}
F_Z(z)&= \frac{1}{{{2^L}\Gamma (L)}}{e^{ - \xi}}\int\limits_0^z {{x^{L - 1}}{e^{ - \frac{x}{2}}}\sum\limits_{i = 0}^\infty  {\frac{{{{(\frac{1}{2}\xi x)}^i}}}{{{{(L)}_i}i!}}} dz} \notag \\
 &= {e^{ - \xi }}\sum\limits_{i = 0}^\infty  {\frac{{(\xi )}^i}{i!}\frac{{\gamma (L + i,\frac{z}{2})}}{{\Gamma (L + i)}}}.
\end{align}
By substituting $z=\frac{x}{\Psi_{gNLoS}/2}$ into (\ref{eqn:CDF_Z}), we can obtain the CDF of $G$ as (\ref{eqn:CDF_CC}). Thus we complete the proof.


\bibliographystyle{ieeetr}
\bibliography{RIS_HARQ}

\begin{thebibliography}{10}

\bibitem{hu2018beyond}
S.~Hu, F.~Rusek, and O.~Edfors, ``Beyond massive {MIMO}: The potential of
  positioning with large intelligent surfaces,'' {\em IEEE Trans. on Signal
  Process.}, vol.~66, pp.~1761--1774, Apr. 2018.

\bibitem{wu2020towards}
Q.~Wu and R.~Zhang, ``Towards smart and reconfigurable environment:
  {Intelligent} reflecting surface aided wireless network,'' {\em IEEE Commun
  Mag.}, vol.~58, pp.~106--112, Jan. 2020.

\bibitem{basar2020reconfigurable}
E.~Basar, ``Reconfigurable intelligent surface-based index modulation: {A} new
  beyond {MIMO} paradigm for {6G},'' {\em IEEE Trans. Commun.}, vol.~68,
  pp.~3187--3196, May 2020.

\bibitem{ElMossallamy2020}
M.~A. {El Mossallamy}, H.~{Zhang}, L.~{Song}, K.~G. {Seddik}, Z.~{Han}, and
  G.~Y. {Li}, ``Reconfigurable intelligent surfaces for wireless
  communications: Principles, challenges, and opportunities,'' {\em IEEE Trans.
  Cogn. Commun. Netw.}, vol.~6, pp.~990--1002, Sept. 2020.

\bibitem{wu2019intelligent}
Q.~Wu and R.~Zhang, ``Intelligent reflecting surface enhanced wireless network
  via joint active and passive beamforming,'' {\em IEEE Trans. Wireless
  Commun.}, vol.~18, pp.~5394--5409, Nov. 2019.

\bibitem{kammoun2020asymptotic}
A.~Kammoun, A.~Chaaban, M.~Debbah, M.-S. Alouini, {\em et~al.}, ``Asymptotic
  max-min {SINR} analysis of reconfigurable intelligent surface assisted {MISO}
  systems,'' {\em IEEE Trans. Wireless Commun.}, vol.~PP, no.~99, pp.~1--1,
  2020.

\bibitem{boulogeorgos2020performance}
A.-A.~A. Boulogeorgos and A.~Alexiou, ``Performance analysis of reconfigurable
  intelligent surface-assisted wireless systems and comparison with relaying,''
  {\em IEEE Access}, vol.~8, pp.~94463--94483, Jun. 2020.

\bibitem{li2020weighted}
Z.~Li, M.~Hua, Q.~Wang, and Q.~Song, ``Weighted sum-rate maximization for
  multi-{IRS} aided cooperative transmission,'' {\em IEEE Wireless Commun.
  lett.}, vol.~PP, no.~99, pp.~1--1, 2020.

\bibitem{Atapattu2020reconfigurable}
S.~{Atapattu}, R.~{Fan}, P.~{Dharmawansa}, G.~{Wang}, J.~{Evans}, and T.~A.
  {Tsiftsis}, ``Reconfigurable intelligent surface assisted {Two-Way}
  communications: Performance analysis and optimization,'' {\em IEEE Trans.
  Wireless Commun.}, vol.~PP, no.~99, pp.~1--1, 2020.

\bibitem{ji2018ultra}
H.~Ji, S.~Park, J.~Yeo, Y.~Kim, J.~Lee, and B.~Shim, ``Ultra-reliable and
  low-latency communications in {5G} downlink: Physical layer aspects,'' {\em
  IEEE Wireless Commun.}, vol.~25, pp.~124--130, Jun. 2018.

\bibitem{shi2017asymptotic}
Z.~Shi, S.~Ma, G.~Yang, K.-W. Tam, and M.~Xia, ``Asymptotic outage analysis of
  {HARQ-IR} over time-correlated {Nakagami-$m$} fading channels,'' {\em IEEE
  Trans. Wireless Commun.}, vol.~16, pp.~6119--6134, Sept. 2017.

\bibitem{shi2018energy}
Z.~Shi, S.~Ma, G.~Yang, and M.-S. Alouini, ``Energy-efficient optimization for
  {HARQ} schemes over time-correlated fading channels,'' {\em IEEE Trans. Veh.
  Technol.}, vol.~67, pp.~4939--4953, Jun. 2018.

\bibitem{shi2019achievable}
Z.~Shi, C.~Zhang, Y.~Fu, H.~Wang, G.~Yang, and S.~Ma, ``Achievable diversity
  order of {HARQ-Aided} downlink {NOMA} systems,'' {\em IEEE Trans. Veh.
  Technol.}, vol.~69, pp.~471--487, Jan. 2020.

\bibitem{Caire2001the}
G.~{Caire} and D.~{Tuninetti}, ``The throughput of hybrid-{ARQ} protocols for
  the gaussian collision channel,'' {\em IEEE Trans. Inf. Theory}, vol.~47,
  pp.~1971--1988, Jul. 2001.

\bibitem{Wang2020outage}
H.~{Wang}, Z.~{Shi}, Y.~{Fu}, and R.~{Song}, ``Outage performance for
  {NOMA-Aided} small cell networks with {HARQ},'' {\em IEEE Wireless Commun.
  Lett.}, vol.~PP, no.~99, pp.~1--1, 2020.

\bibitem{chelli2018throughput}
A.~Chelli, E.~Zedini, M.-S. Alouini, M.~P{\"a}tzold, and I.~Balasingham,
  ``Throughput and delay analysis of harq with code combining over double
  rayleigh fading channels,'' {\em IEEE Trans. Veh. Technol.}, vol.~67,
  pp.~4233--4247, May 2018.

\bibitem{Shi2015analysis}
Z.~{Shi}, H.~{Ding}, S.~{Ma}, and K.~{Tam}, ``Analysis of {HARQ-IR} over
  time-correlated rayleigh fading channels,'' {\em IEEE Trans. Wireless
  Commun.}, vol.~14, pp.~7096--7109, Dec. 2015.

\bibitem{xiong1997performance}
F.~Xiong and S.~Bhatmuley, ``Performance of {MHPM} in {Rician} and {Rayleigh}
  fading mobile channels,'' {\em IEEE Trans. Commun.}, vol.~45, pp.~279--283,
  Mar. 1997.

\bibitem{hu2018user}
S.~Hu, K.~Chitti, F.~Rusek, and O.~Edfors, ``User assignment with distributed
  large intelligent surface ({LIS}) systems,'' in {\em Proc. IEEE Annu. Int.
  Symp. Pers., Indoor, Mobile Radio Commun. (PIMRC)}, (Bologna, Italy),
  pp.~1--6, Sept. 2018.

\bibitem{yang2020asymptotic}
G.~Yang, H.~Zhang, Z.~Shi, S.~Ma, and H.~Wang, ``Asymptotic outage analysis of
  spatially correlated rayleigh {MIMO} channels,'' {\em IEEE Trans.
  Broadcast.}, vol.~PP, no.~99, pp.~1--1, 2020.

\bibitem{zhang2019analysis}
Z.~Zhang, Y.~Cui, F.~Yang, and L.~Ding, ``Analysis and optimization of outage
  probability in multi-intelligent reflecting surface-assisted systems,'' {\em
  arXiv:1909.02193}, 2019.

\bibitem{gradshteyn2014table}
I.~S. Gradshteyn and I.~M. Ryzhik, {\em Table of Integrals, Series, and
  Products}.
\newblock Academic press, 2014.

\bibitem{kudathanthirige2020performance}
D.~{Kudathanthirige}, D.~{Gunasinghe}, and G.~{Amarasuriya}, ``Performance
  analysis of intelligent reflective surfaces for wireless communication,'' in
  {\em Proc. 2020 IEEE International Conference on Communications (ICC)},
  (Dublin, Ireland), pp.~1--6, Jul. 2020.

\bibitem{muirhead2009aspects}
R.~J. Muirhead, {\em Aspects of Multivariate Statistical Theory}, vol.~197.
\newblock John Wiley \& Sons, 2009.

\end{thebibliography}

\end{document}